\documentstyle[preprint,prd,aps,epsf]{revtex}
\tightenlines
\begin{document}
\draft

\title{Complex Visibilities of \\Cosmic Microwave Background Anisotropies}

\author{Kin-Wang Ng\footnote
{Current address: Canadian Institute for Theoretical Astrophysics, 
60 St. George St., Toronto, ON M5S 3H8, Canada.}}
\address{Institute of Physics \& Institute of Astronomy and Astrophysics,\\
Academia Sinica, Taipei, Taiwan, R.O.C.}

\date{September 2000}
\maketitle

\begin{abstract}
We study the complex visibilities of the cosmic microwave background 
anisotropies that are observables in interferometric observations of
the cosmic microwave background, using the multipole expansion methods
commonly adopted in analyzing single-dish experiments. This allows us
to recover the properties of the visibilities that is obscured in
the flat-sky approximation. Discussions of the window function, multipole
resolution, instrumental noise, pixelization, and polarization are given.  
\end{abstract}

\vspace{2pc}
\pacs{PACS number(s): 98.70.Vc, 98.80.Es}

\section{Introduction}

The detection of the large-angle temperature anisotropy of the 
Cosmic Microwave Background (CMB) by the 
{\it COBE} DMR experiment~\cite{smo} provided important evidence of
large-scale spacetime inhomogenities. Since then, many 
CMB measurements have reported detections or upper
limits of the CMB anisotropy power spectrum
over a wide range of scales~\cite{pag}.  
Recently, BOOMERANG~\cite{boo} and MAXIMA~\cite{max} data 
have revealed the structure of the first Doppler peak 
which arises from acoustic oscillations of the baryon-photon plasma on
the last scattering surface.
Future ground-based and balloon-borne experiments, 
and upcoming NASA MAP satellite will unveil detailed features of the CMB 
anisotropy, thus allowing one to determine to
a high precision a number of cosmological parameters~\cite{jun}.

The CMB is linearly polarized when the
anisotropic radiation is scattered with free electrons 
near the last scattering surface~\cite{ree}. 
The degree of polarization is about one-tenth of the
temperature anisotropy at sub-degree scales~\cite{bon}.
The CMB polarization contains a wealth of information about the early 
Universe as well. It provides a sensitive test of the reionization history
as well as the presence of non-scalar metric perturbations, and improves
the accuracy in determining the cosmological parameters~\cite{zss}.
So far, the current upper limit on the CMB linear
polarization is $16\mu K$~\cite{net}. A handful of new experiments,
adopting low-noise receivers as well as long integration time per
pixel, are underway or being planned~\cite{sta}.

CMB experiments are commonly single-dish chopping instruments,
whose scanning strategy and data analysis procedure are well developed.
In the last decade, interferometers were introduced to study the microwave sky.
Recent advancement in low-noise, broadband, GHz amplifiers, 
in addition to mature synthesis imaging techniques, 
has made interferometry a particularly attractive technique for 
detecting CMB anisotropies.  An interferometric array
is intrinsically a high-resolution instrument well-suited 
for observing small-scale intensity fluctuations, while being flexible in 
coverage of a wide range of angular scales with the resolution and 
sensitivity determined by the aperture of each element of the array
and the baselines formed by the array elements.
A desirable feature of the interferometer for CMB observation is 
that it directly measures the power spectrum of the sky.
In addition, many systematic problems that are inherent in single-dish
experiments, such as ground and near field atmospheric pickup, and
spurious polarization signal, can be reduced or avoided in interferometry. 
A brief account of the interferometic CMB observations can be found 
in Ref.~\cite{white}. Currently, several new interferometers including 
the Very Small Array (VSA)~\cite{vsa}, Degree Angular Scale
Interferometer (DASI)~\cite{dasi}, Cosmic Background Imager 
(CBI)~\cite{cbi}, Array for Microwave Background Anisotropy
(AMiBA)~\cite{amiba} are being built, 
with sensitivity about a few $\mu K$ per beam. 
As for the polarization capability, the AMiBA will have full polarizations,
whereas the VSA, DASI, and CBI will not be polarization sensitive initially. 

The output of an interferometer is the visibility that is the Fourier 
transform of the intensity fluctuations on the sky. 
In this paper, we will study the CMB anisotropies in 
the visibility space. There are several papers dealing with 
CMB anisotropy data from interferometers~\cite{mar,tim,ravi,hob}, 
and making explicit contact of the visibility with the 
angular power spectrum in $l$-space that is frequently used in single-dish
experiments~\cite{white,hob2,ravi2,white2}. All of them have treated the CMB
sky as flat due to the fact that typically a small field on the sky is 
viewed by the interferometer. This turns the analysis into a familiar
two-dimensional Fourier transform problem. 
Here rather than assuming the flat-sky approximation, 
we will perform a full-sky analysis of the visibility. 
Although we then have to give
up the relatively simple Fourier-transform formalism, the bonus is that 
the angular power spectrum can be directly transferred to the visibility space.
Hence, the statistics of the CMB visibilities 
is straightforward induced by the Gaussian variables in $l$-space.

The paper is organized as follows. In Sec.~\ref{cmb}, a brief
account of the building block of CMB anisotropies is given. 
We introduce the interferometric observation in Sec.~\ref{int}, 
and present the calculations of the CMB visibilities in Sec.~\ref{cvis}. 
The resolution in $l$-space is discussed in Sec.~\ref{res}, 
the instrumental noise and pixelization is introduced in Sec.~\ref{noise}, 
and the CMB polarization is treated in Sec.~\ref{poln}.
Estimation of signal to noise ratios for interferometric CMB measurements
are made in Sec.~\ref{snr}. Sec.~\ref{con} is our conclusions and discussions.

\section{CMB Anisotropies}
\label{cmb}

Before we study the interferometric observation of CMB anisotropies,
let us review the basic ingredients that are based upon the multipole 
expansion of the Gaussian random fields and commonly used 
in analyzing single-dish experiments

Polarized emission is conventionally described in terms of the four Stokes
parameters $(I,Q,U,V)$, where $I$ is the intensity, $Q$ and $U$ represent 
the linear polarization, and $V$ describes the circular polarization.
Since circular polarization cannot be generated by Thomson scattering alone, 
the parameter $V$ decouples from the other components and will not
be considered. Let us define $T$ be the temperature fluctuation about the mean,
then the CMB anisotropies are completely described by $(T,Q,U)$, where
each parameter is a function of the pointing direction $\hat e$
on the celestial sphere.

Considering the CMB as Gaussian random fields, we can 
expand the Stokes parameters as~\cite{zald,kamion}
\begin{eqnarray}
T(\hat e)&=&\sum_{lm}a_{T,lm}Y_{lm}(\hat e), \nonumber \\
(Q-iU)(\hat e)&=&\sum_{lm}a_{2,lm}\:_{2}Y_{lm}(\hat e), \nonumber \\
(Q+iU)(\hat e)&=&\sum_{lm}a_{-2,lm}\:_{-2}Y_{lm}(\hat e),
\label{expand}
\end{eqnarray}
where $a_{T,lm}$ and $a_{\pm 2,lm}$ are Gaussian random variables, and
$\:_{\pm 2}Y_{lm}$ are spin-2 spherical harmonics.
Details of the spin-weighted spherical harmonics and their properties 
can be found in Appendix~A. 

Isotropy in the mean guarantees the following ensemble averages:
\begin{eqnarray}
\left<a^{*}_{T,l'm'}a_{T,lm}\right>&=&C_{Tl}\delta_{l'l}\delta_{m'm},
\nonumber \\
\left<a^{*}_{2,l'm'}a_{2,lm}\right>&=&(C_{El}+C_{Bl})\delta_{l'l}
\delta_{m'm}, \nonumber \\ 
\left<a^{*}_{2,l'm'}a_{-2,lm}\right>&=&(C_{El}-C_{Bl})
\delta_{l'l}\delta_{m'm}, \nonumber \\
\left<a^{*}_{T,l'm'}a_{2,lm}\right>&=&-C_{Cl}\delta_{l'l}\delta_{m'm},
\label{CMBaa}
\end{eqnarray}
where $C_{Tl}$, $C_{El}$, $C_{Bl}$, and $C_{Cl}$ 
are respectively the anisotropy, E-polarization, B-polarization,
and T-E cross correlation angular power spectra.

Consider two pointings $\hat e_1$ and $\hat e_2$ 
on the celestial sphere.  
Using the generalized addition theorem~(\ref{add}) and Eq.~(\ref{CMBaa}), 
we obtain the correlation functions~\cite{ngliu},
\begin{eqnarray}
&&\left<T^{*}(\hat e_1)T(\hat e_2)\right>
  =\sum_l\frac{2l+1}{4\pi}C_{Tl} P_l(\cos\beta), \nonumber \\
&&\left<T^{*}(\hat e_1)[Q(\hat e_2)+iU(\hat e_2)]\right>
  =-\sum_l\frac{2l+1}{4\pi}\sqrt{\frac{(l-2)!}{(l+2)!}}C_{Cl}
   P^2_l(\cos\beta) e^{-2i\alpha_2}, \nonumber \\
&&\left<[Q(\hat e_1)+iU(\hat e_1)]^*[Q(\hat e_2)+iU(\hat e_2)]\right>
  =\sum_{l}\sqrt{\frac{2l+1}{4\pi}}(C_{El}+C_{Bl})
   \:_{2}Y_{l-2}(\beta,0)e^{-2i(\alpha_2-\alpha_1)}, \nonumber \\
&&\left<[Q(\hat e_1)-iU(\hat e_1)]^*[Q(\hat e_2)+iU(\hat e_2)]\right>
  =\sum_{l}\sqrt{\frac{2l+1}{4\pi}}(C_{El}-C_{Bl})
   \:_{2}Y_{l2}(\beta,0)e^{-2i(\alpha_2+\alpha_1)}, 
\label{cfe}
\end{eqnarray}
where $\beta$, $\alpha_1$, and $\alpha_2$ are the angles defined in Appendix~A.
Therefore, the statistics of the CMB anisotropy and polarization is fully
described by the four independent power spectra
or their corresponding correlation functions. The details about the evaluation
of the power spectra can be found in Refs.~\cite{zald,kamion}.

In realistic CMB observations, due to the finite beam size of the antenna
and the beam switching mechanism, a measurement is actually a convolution
of the antenna response and the Stokes parameters. 
This can be accounted by a mapping of the spherical
harmonics in Eq.~(\ref{expand}),
\begin{equation}
\:_{s}Y_{lm}(\hat e)\to (-1)^s \:_{s}W_{lm}^{1\over 2} \:_{s}Y_{lm}(\hat e),
\end{equation}
where $\:_{s}W_{lm}$ is the window function. For a simple
single-dish experiment with Gaussian beamwidth $\sigma_b$, 
it was found that~\cite{ngliu}
\begin{equation}
\:_{s}W_{lm} \simeq \exp\left[-\left(l(l+1)-s^2\right)\sigma_b^2\right].
\label{swinf}
\end{equation}

\section{Interferometric Observation}
\label{int}

In contrast to single-dish experiments that measure or differentiate
the signals in individual dishes, an interferometer measures
correlation of the signals from different pairs of the array elements.
Let us consider a two-element instrument and a monochromatic electromagnetic
source. The output, called the {\it complex visbility}, of the interferometer
is the time-averaged correlation of the electric field $E$ measured by two
antennae pointing in the same direction to the sky but at two different
locations~\cite{thom}:
\begin{equation}
V(\vec u) = \langle E_1 E_2^* \rangle
          = \int d{\hat e} A(\hat e,\hat e_0) I(\hat e) 
                             e^{2\pi i{\vec u}\cdot {\hat e}},
\label{fullv}
\end{equation}
where $\vec u$ is the separation vector (baseline) of the two antennae 
measured in units of the observation wavelength,
$A$ denotes the primary beam with the phase tracking center 
pointing along the unit vector $\hat e_0$, 
and $I$ is the intensity of the source. 
Note that $\vec u$ is generally a three-dimensional vector, 
and that $V^*(\vec u)=V(-\vec u)$ since $A$ and $I$ are real functions. 

In the following, we consider a primary antenna with Gaussian beamwidth 
$\sigma_b$ given by
\begin{equation}
A(\theta,\phi)=e^{-\frac{\theta^2}{2\sigma_b^2}},
\label{beam}
\end{equation}
where $(\theta,\phi)$ are the polar angles with respect to $\hat e_0$.
Antenna theory~\cite{kraus} states that
\begin{equation}
A_{eff} \Omega_A = \lambda^2,\quad{\rm with}\quad
\Omega_A = \int d\Omega A(\theta,\phi),
\end{equation}
where $\lambda$ is the observation wavelength, $\Omega_A$ is the field of view,
and the effective area is the aperature efficiency times 
the physical area of the antenna:
\begin{equation}
A_{eff}= \eta_a A_{phy}.
\end{equation}
Let us assume a circular dish with diameter $D$, then we have
\begin{equation}
\sigma_b=\frac{\lambda}{\pi D}\sqrt{\frac{2}{\eta_a}}.
\label{bw}
\end{equation}

In typical interferometric measurements, we have $\lambda \ll D$, dictating
a small $\Omega_A$. Thus, for a single pointing, it is very good to make
the flat-sky approximation by decomposing
\begin{equation}
\hat e = \hat e_0 + {\bf x} ,\quad{\rm with}\quad {\bf x} \cdot \hat e_0 =0,
\quad{\rm and}\quad \left\vert {\bf x} \right\vert \ll 1,
\end{equation}
meaning that ${\bf x}$ is a two-dimensional vector lying in the plane of the 
sky. Hence, the complex visibilty is reduced to the two-dimensional
Fourier transform of the sky intensity multiplied by the primary beam:
\begin{equation}
V({\bf u}) \equiv V(\vec u) e^{-2\pi i{\vec u}\cdot {\hat e_0}} \simeq
\int d{\bf x} A({\bf x}) I({\bf x}) e^{2\pi i{\bf u}\cdot {\bf x}},
\end{equation}
where ${\bf u}$ is the two-dimensional
projection vector of $\vec u$ in the ${\bf x}$-plane.

In a single-dish experiment, the resolution of the image is limited by the 
finite primary beamwidth. In contrast, the interferometric imaging has 
resolution of the synthesized beamwidth which is determined by the 
sampling of the visibility plane (${\bf u}$-plane), 
where the minimum spacing is the physical size of the dish while the maximum 
spacing is limited by the size of the platform which resides the dishes.
The sampling can be described by a sampling function $S({\bf u})$, which is
zero where no data have been taken. One can then perform the inverse
Fourier transform to form the so-called dirty image,
\begin{equation}
I^D({\bf x}) =\int d{\bf u} V({\bf u}) 
S({\bf u}) e^{-2\pi i{\bf u}\cdot {\bf x}}.
\end{equation}
Using the convolution theorem, $I^D=AI\star B$, where $B({\bf x})$ is the 
synthesized beam that is the Fourier transform of $S({\bf u})$.
So, in the flat-sky approximation the calculation is simplified 
into a two-dimensional Fourier transform problem. 
However, the CMB is a large source. Although it is valid to treat the
sky as flat due to the small field of view of the primary beam in a single 
pointing, a larger scale CMB image requires multiple pointings on the sky. 
In addition, after replacing locally the sphere by a plane, the global
property of the CMB field is obscured although there is a direct
link between the angular power spectrum and the power spectrum in the plane.
Therefore, in the following we will begin with the 
general form for the complex visibility in Eq.~(\ref{fullv}).

\section{Complex Visibility of CMB Anisotropy}
\label{cvis}

The CMB brightness fluctuation is related to the temperature fluctuation by
\begin{equation}
\Delta B_\nu = \frac{\partial B_\nu}{\partial T} \Delta T, 
\quad {\rm with} \quad
\frac{\partial B_\nu}{\partial T} \simeq 99.27 f(x)
{\rm Jy\;sr^{-1}}\;\mu{\rm K^{-1}},
\end{equation}
where $B_\nu$ is the Planck function of the photon frequency $\nu$, and
\begin{equation}
f(x)= \frac{x^4 e^x}{(e^x-1)^2},\quad{\rm where}\quad
x\simeq 1.76 \left(\frac{\nu}{100{\rm GHz}}\right).
\label{f(x)}
\end{equation}
We write $\Delta T\equiv T$ and then do the 
anisotropy angular power spectrum expansion~(\ref{expand}).

\subsection{Infinite Resolution Limit}
\label{inf}

Let us concentrate on the power spectrum and neglect the effect of the
primary beam by taking for now the infinite resolution limit that $A=1$,
then from Eq.~(\ref{fullv}) the complex visibilty of the CMB anisotropy 
is written as
\begin{equation}
V(\vec u) = \frac{\partial B_\nu}{\partial T} V_T(\vec u), 
\quad{\rm where}\quad
V_T(\vec u) = \int d\hat e  T(\hat e) e^{2\pi i{\vec u}\cdot {\hat e}}.
\label{VT}
\end{equation}
Also expanding the phase factor in terms of spherical waves,
\begin{equation}
e^{2\pi i{\vec u}\cdot {\hat e}}
=4\pi \sum_{lm} i^l j_l(2\pi u) Y^*_{lm}(\hat e) Y_{lm}(\hat u),
\label{expon}
\end{equation}
where $j_l$ is the spherical Bessel junction and
$u=\left\vert \vec u \right\vert$, we obtain
\begin{equation}
V_T(\vec u) = 4\pi \sum_{lm} a_{T,lm} i^l j_l(2\pi u) Y_{lm}(\hat u).
\label{VTl}
\end{equation}
For a fixed length of the baseline, Eq.~(\ref{VTl})
is analogous to a measurement of the temperature anisotropy $T(\hat e)$ 
in real space with a window function. 
So, the statistics of $V_T$ on a sphere of radius $u$ in the visibility space 
is completely described by the $a_{T,lm}$ and the positively definite function 
\begin{equation}
I_l(u)=16\pi^2 j_l^2(2\pi u).
\label{intf}
\end{equation}
We will call this function the {\it interference function}. 
Estimation of the $C_{Tl}$'s can be made from the visibility data 
in the same way as one does for an anisotropy sky map, with the 
sample variances determined by the coverage of the visibility sphere. 

Using Eq.~(\ref{CMBaa}), the two-point correlation function is given by 
\begin{equation}
 \langle V^*_T(\vec u_1) V_T(\vec u_2) \rangle =
    \frac{1}{4\pi} \sum_l (2l+1) C_{Tl} I_l(u) P_l(\cos\theta),
\label{cf}
\end{equation}
where $P_l$ is the Legendre polynomial and 
$\cos\theta={\hat u_1} \cdot {\hat u_2}$. 
Hence, the rms temperature anisotropy in a given visibility is 
\begin{equation}
V_{T{rms}}^2 \equiv \langle V^*_T(\vec u) V_T(\vec u) \rangle
= \frac{1}{4\pi} \sum_l (2l+1) C_{Tl} I_l(u).
\label{power}
\end{equation}
On the other hand, based on the flat-sky approximation, 
the authors in Ref.~\cite{white} have obtained a similar autocorrelation 
function but with a different interference function,
\begin{equation}
I^{fs}_l(u)=\frac{2}{u} J_{2l+1}(4\pi u),
\label{fsintf}
\end{equation}
where $J_{2l+1}$ is the Bessel function. A comparison between these
two interference functions is made in Fig.~\ref{fig1} with $2\pi u=200$.
Both interference functions have a high peak at $l\sim 200$ and drop rapidly
as $l>200$. While $I_l$ is positively definite, $I^{fs}_l$ is oscillatory 
about the zero and has a lower peak. As such the latter would underestimate 
the rms fluctuation in a given visibility. 
For instance, assuming a flat anisotropy power
spectrum, we find that the flat-sky rms visibility is about one-fourth
of that in Eq.~(\ref{power}).

\subsection{Finite Primary Beam}

In practice, the instrument is limited by a finite primary beamwidth
given by the aperture function.
Inserting the aperture function with tracking
direction $\hat e$ in Eq.~(\ref{VT}), we have
\begin{equation}
V_T(\vec u,\hat e) =\int d{\hat e'} A(\hat e',\hat e) T(\hat e') 
                             e^{2\pi i{\vec u}\cdot {\hat e'}}.
\label{VTb}
\end{equation}
Doing the expansions~(\ref{expand}) and (\ref{expon}), assuming a
Gaussian beam~(\ref{beam}), and using Eq.~(\ref{add2}), 
Eq.~(\ref{VTb}) becomes
\begin{eqnarray}
V_T(\vec u,\hat e) = && 8\pi^2 \sum_{l_1 m_1} \sum_{l_2 m_2}
   a_{T, l_1 m_1} i^{l_2} j_{l_2}(2\pi u) Y_{l_2 m_2}(\hat u)
   \sqrt{\frac{4\pi}{2l_1+1}}\sqrt{\frac{4\pi}{2l_2+1}} \times \nonumber \\
   && \sum_{m} \:_{-m}Y^*_{l_2 m_2}(\hat e) \:_{-m}Y_{l_1 m_1} (\hat e)
   W^{1\over2}(l_1,l_2,m),   
\label{VTW}
\end{eqnarray}
where the square root of the window function is
\begin{equation}
W^{1\over2}(l_1,l_2,m)=\int_0^\pi 
                       d\beta~\beta~e^{-\frac{\beta^2}{2\sigma_b^2}}
                       Y^*_{l_2 m}(\beta,0) Y_{l_1 m}(\beta,0),
\label{winf}
\end{equation}
and $\:_{s}Y_{lm}$ is the spin-$s$ spherical harmonics (see Appendix~A).
Eq.~(\ref{VTW}) is the general result for
the complex visibility of the CMB observed by an interferometer with
two identical Gaussian primary beams separated by a baseline $\vec u$.
However, it is difficult to get any useful information from its present form. 
In the following, we will pursue two configurations 
for which Eq.~(\ref{VTW}) can be boiled down to useful forms. 
The first one is the ``close-packed'' configuration with
the dishes almost in touching, i.e., $u\lambda \sim D$.
This configuration is commonly adopted in 
CMB observations in order to maximize the size of the dish to
obtain optimal sensitivity for signal detection. 
The second is the ``widespread'' configuration in which the length of the
baseline is much longer than the diameter of the dish, i.e.,
$u\lambda \gg D$. This configuration is needed when one wants to 
resolve the fine structure of an image.

\subsubsection{Close-packed}

Let us start by analyzing the interference function, which is composed of
the spherical Bessel function $j_l(2\pi u)$.
Typically, the baseline $u\gg 1$. For large argument, $j_l$ has a sharp
peak at $l\sim 2\pi u$. The asymptotic expansion in the peaked region is
given by~\cite{abr}
\begin{equation}
j_l (2\pi u) \sim {1\over2} u^{-\frac{1}{2}} 2^{1\over 3} 
                  (l+{1\over2})^{-\frac{1}{3}} {\rm Ai}(-2^{1\over 3}y),\quad
                  {\rm where}\quad 
                  2\pi u=(l+{1\over2}) + y (l+{1\over2})^{\frac{1}{3}},
\label{asymj}
\end{equation}
and Ai($z$) is the Airy function.
Thus, for a fixed $u$, the location and width of the peak are approximately
given by
\begin{equation}
l_{pk} + 0.8 l_{pk}^{\frac{1}{3}} =2\pi u,
\quad{\rm and}\quad
\Delta l_I \sim l_{pk}^{\frac{1}{3}}.
\end{equation} 
In addition, the height of the peak is given by
\begin{equation}
j_{l_{pk}} (2\pi u) \simeq \sqrt{\pi} 2^{-{1\over 6}} l_{pk}^{-{5\over6}}
{\rm Ai}(-1);
\quad {\rm Ai}(-1) \simeq 0.54.
\end{equation} 
In Fig.~\ref{fig1}, we have used the asymptotic expansion~(\ref{asymj})
to reproduce the interference function in Eq.~(\ref{intf}). It shows that
the asymptote is a fairly good approximation. 

In the close-packed configuration, we denotes the minimum spacing
$l_{min} = 2\pi u$, where from Eq.~(\ref{bw})
\begin{equation}
l_{min} \sim 2\pi \frac{D}{\lambda}=
\frac{2}{\sigma_b}\sqrt{\frac{2}{\eta_a}} \gg 1.
\end{equation}
Because the dominant contribution in the $l_2$-summation 
in Eq.~(\ref{VTW}) comes from $l_2\lesssim l_{min}$, we can approximate the
associate Legendre polynomial in the integrand~(\ref{winf}) by~\cite{gra},
\begin{equation}
P^{-m}_{l_2} (\cos\beta) \simeq l_2^{-m} J_m(l_2\beta).
\end{equation}
Hence, the integral~(\ref{winf}) can be evaluated analytically 
for certain limiting values of $l_1$. 
The integration is detailed in Appendix~B.
Essentially, the window function can be approximated by
\begin{equation}
W^{1\over2}(l_1,l_2,m) \simeq 
\left[\frac{2l_1+1}{4\pi} \frac{(l_1+m)!}{(l_1-m)!} \right]^{1\over2}
\left[\frac{2l_2+1}{4\pi} \frac{(l_2+m)!}{(l_2-m)!} \right]^{1\over2} 
\frac{1}{(l_1 l_2)^{m+{1\over2}}} \frac{\sigma_b}{\sqrt{2\pi}}
e^{-{1\over2}(l_1-l_2)^2 \sigma_b^2}.
\label{cpwf}
\end{equation}
Note that the window function has a Gaussian width, 
$\Delta l_W \sim l_{min}$, and that it
decreases with $m$ rapidly. The latter enables us to retain only $m=0$
in the $m$-summation in Eq.~(\ref{VTW}), which then becomes
\begin{equation}
V_T(\vec u,\hat e) \simeq \sqrt{2\pi} \sigma_b \sum_{l_1 m_1} \sum_{l_2}
   a_{T, l_1 m_1}  Y_{l_1 m_1} (\hat e) i^{l_2} (2l_2+1) j_{l_2}(2\pi u)
   P_{l_2} (\hat e \cdot \hat u)
   \frac{1}{\sqrt{l_1 l_2}} e^{-{1\over2}(l_1-l_2)^2 \sigma_b^2}.  
\label{cpv1}
\end{equation}
In the close-packed configuration, we usually have the 
condition that $\Delta l_W \gg \Delta l_I$. As such, 
as compared to the window function, the interference
function can be treated as a delta function at $l_2=l_{pk}$,
where $l_{pk} + 0.8 l_{pk}^{1/3} = l_{min}$. For $l_{min} \gg 1$,
we have $l_{pk} \sim l_{min}$. Hence, we approximate 
\begin{equation}
\frac{1}{\sqrt{l_1 l_2}} e^{-{1\over2}(l_1-l_2)^2 \sigma_b^2}
\sim
\frac{1}{l_{min}} e^{-{1\over2}(l_1-l_{min})^2 \sigma_b^2}
\end{equation}
in Eq.~(\ref{cpv1}), which then turns into the final form,
\begin{equation}
V_T(\hat e) \simeq e^{i l_{min}{\hat u}\cdot {\hat e}}
                   \frac{\sqrt{2\pi} \sigma_b}{l_{min}}
                   \sum_{lm} a_{T,lm} Y_{lm}(\hat e) W^{1\over2}_l, 
\label{cpv2}
\end{equation}
where the window function is
\begin{equation}
W_l = e^{-(l-l_{min})^2 \sigma_b^2}.
\end{equation}
The result~(\ref{cpv2}) shows that the baseline vector appears 
only in the irrelevant phase factor, and thus that 
the close-packed interferometric beam is similar to a single-beam antenna 
undergoing chopping and wobbling. In fact, this similarity has been pointed out 
in an early CMB interferometric measurement~\cite{tim}. 
So, when we attempt to construct the two-point correlation function 
similar to  Eq.~(\ref{cf}) with $|\vec u_1| \sim |\vec u_2| \sim D/\lambda$,
we obtain a pure phase function which does not contain any useful information. 
The reason is simply that the correlation over the domain in the
$\vec u$-space spanned by $\vec u_1$ and $\vec u_2$, 
whose size is still comparable to the size of the dish, is almost a constant.
In light of this, in Eq.~(\ref{cpv2})
we have replaced $V_T(\vec u,\hat e)$ by 
$V_T(\hat e)$ to reflect that it is more legitimate to deal with 
the visibility in $\hat e$-space than in $\vec u$-space.
Therefore, for the close-packed configuration with a fixed baseline, 
we should sample $V_T(\hat e)$ at different parts of the sky by 
repointing the entire telescope, 
and analyze the data in the same way as in the single-dish observation.

\subsubsection{Widespread}
\label{wsp}

In this case, the dominant contribution in the $l_2$-summation 
in Eq.~(\ref{VTW}) comes from the range,
$l_{min} \ll l_2 \lesssim 2\pi u$, where we can approximate the
associate Legendre polynomial in the integrand~(\ref{winf}) by~\cite{gra},
\begin{equation}
P^{m}_{l_2} (\cos\beta) \simeq \sqrt{2\over\pi}
\frac{\Gamma(l_2+m+1)}{\Gamma(l_2+{3\over2})}
\sin^{-{1\over2}}\beta~ 
\cos\left[(l_2+{1\over2})\beta + (m-{1\over2}){\pi\over2}\right].
\end{equation}
Hence, the window function~(\ref{winf}) can be approximately evaluated as
\begin{eqnarray}
W^{1\over2}(l_1,l_2,m) \simeq && 
\left[\frac{2l_1+1}{4\pi}~ 
\frac{\Gamma(l_1+m+1)\Gamma(l_1-m+1)}{\Gamma^2(l_1+{3\over2})}
\right]^{1\over2} \times \nonumber \\
&& \left[\frac{2l_2+1}{4\pi}~
\frac{\Gamma(l_2+m+1)\Gamma(l_2-m+1)}{\Gamma^2(l_2+{3\over2})}
\right]^{1\over2} 
\frac{\sigma_b}{\sqrt{2\pi}} e^{-{1\over2}(l_1-l_2)^2 \sigma_b^2}.
\label{wswf}
\end{eqnarray}
We have detailed the integration in Appendix~B.
Note that the window function~(\ref{wswf}) has a very narrow Gaussian width
$\Delta l_W \sim l_{min}$, when compared to the baseline length $\sim 2\pi u$.
As such, we can make $l_1 \sim l_2$. Then, using the asymptotic formula
\begin{equation}
\frac{\Gamma(l\pm m+1)}{\Gamma(l+{3\over2})}
\sim \left(l+{3\over2}\right)^{\pm m-{1\over2}}
\quad{\rm for}\quad l\gg 1,
\end{equation}
we can further approximate the window function as
\begin{equation}
W^{1\over2}(l_1,l_2,m) \simeq
(2\pi)^{-{3\over2}} \sigma_b e^{-{1\over2}(l_1-l_2)^2 \sigma_b^2},
\end{equation}
which is independent of $m$ in the asymptotic limit. 
Furthermore, when the condition 
$\Delta l_W \ll \Delta l_I$ is satisfied, the window function can be treated 
as a delta function, 
\begin{equation}
W^{1\over2}(l_1,l_2,m) \simeq
(2\pi)^{-{3\over2}} \sigma_b \delta_{l_1 l_2}.
\label{wsdf}
\end{equation}
Plugging Eqs.~(\ref{wsdf}) and (\ref{yy}) in Eq.~(\ref{VTW}), we finally obtain
\begin{equation}
V_T(\vec u) \simeq \sqrt{8\pi} \sigma_b
             \sum_{lm} a_{T,lm} i^l j_l(2\pi u) Y_{lm}(\hat u),
\label{VTl2}
\end{equation}
which does not depend on the tracking direction. The structure of this equation
is similar to Eq.~(\ref{VTl}), and thus the discussions following
Eq.~(\ref{VTl}) in Sec.~\ref{inf} are equally applied to Eq.~(\ref{VTl2}).

Although the visibility~(\ref{VTl2}) is reduced by the finite beam size, 
it enables us to make independent measurements of the visibility by 
pointing the telescope to different uncorrelated patches of the sky. 
Then, we can extract the $C_{Tl}$'s from the visibility data on
the u-sphere for each pointing. So, if we have $N$ uncorrelated 
pointings and a full u-sphere coverage for each pointing, 
the cosmic variance of the estimation of $C_{Tl}$'s will be 
reduced by a factor of $\sqrt N$. It is interesting to note that
the inevitable cosmic variance in single-dish experiments due to
the existence of a single universe can be in principle circumvented in the 
interferometry. 
However, the widespread condition requires $(2\pi u)^{1/3} \gg 1$, meaning 
a very high resolution scale at which the primary
CMB anisotropies have very low power.

\section{Increasing Resolution}
\label{res}

We have learned in the previous section that the resolution which we have
in $l$-space for a single pointing of the close-packed interferometer
is equal to the size of the primary beam. However, we can
increase the $l$-resolution by combining several contiguous pointings of 
the telescope. This is analogous to the Fraunhofer diffraction in optics,
in which narrower interference fringes are obtained by using many apertures.
In Ref.~\cite{white}, they have demonstrated a case in which the resulting 
aperture in the $u$-plane can be made much narrower by considering
$N \times N$ pointings on a regular grid. Here we can show the increase
of $l$-resolution by combining various pointings of 
$V_T(\vec u,\hat e)$ in Eq.~(\ref{VTW}). 
The simplest case is to sum over all pointing directions,
\begin{equation}
\int d\Omega V_T(\vec u,\hat e) 
= 8\pi^2 \sigma_b^2 \sum_{lm} a_{T,lm} i^l j_l(2\pi u) Y_{lm}(\hat u),
\label{resol}
\end{equation}
where we have used the orthonormality condition~(\ref{ortho}) to evaluate the
integration. This shows that the $l$-resolution is now determined
by the interference function rather than the window function. It means that 
the resolution in $l$ is increased from $\Delta l_W$ to $\Delta l_I$. 
Since Eq.~(\ref{resol}) is a general result, it can be applied to the
widespread configuration in Sec.~\ref{wsp} as well, 
where the $l$-resolution is already given by $\Delta l_I$.
Thus we conclude that in interferometry
the $l$-resolution can be increased by combining telescope pointings
up to a limit set by the intrinsic interference function.

\section{Instrumental Noise}
\label{noise}

In the single-dish CMB experiment, a pixelized map of the 
CMB smoothed with a Gaussian beam
is created. In each pixel, the signal has a 
contribution from the CMB and from 
the instrumental noise. A convenient way of describing the amount of
instrumental noise is to specify the rms noise per pixel $\sigma_{pix}$, 
which depends on the detector
sensitivity $s$ and the time spent observing each pixel $t_{pix}$:
$\sigma_{pix}=s/\sqrt{t_{pix}}$. The noise in each pixel is 
uncorrelated with that in any other pixel, and is uncorrelated 
with the CMB component.
Let $\Omega_{pix}$ be the solid angle subtended by a pixel. 
Usually, given a total observing time, $t_{pix}$
is directly proportional to $\Omega_{pix}$. Thus, we can define a quantity
$w^{-1}$, the inverse statistical weights per unit solid angle, to measure
the experimental sensitivity independent of pixel size~\cite{kno},
\begin{equation}
w^{-1}=\Omega_{pix} \sigma_{pix}^2.
\end{equation}

In interferometry, the instrumental noise is usually specified by 
the image sensitivity per synthesized beam area~\cite{taylor},
\begin{equation}
\Delta I = \frac{1}{\eta_s \eta_a} \frac{2k_B T_{sys}}{A_{phy}}
           \frac{1}{\sqrt{2N_b N_p}} \frac{1}{\sqrt{\Delta_\nu t_{int}}},
\label{DI}
\end{equation}
where $k_B$ is the Boltzmann constant,
$T_{sys}$ and $\eta_s$ are respectively the system temperature and efficiency,
$N_b$ and $N_p$ are respectively the numbers of baselines and polarizations, 
$\Delta_\nu$ is the bandwidth of observation frequency, $t_{int}$ is the 
integration time, and $\eta_a$ and $A_{phy}$ are defined in Sec.~\ref{int}.
For example, the number of baselines formed by $N_a$ antennae
is $N_b=N_a(N_a-1)/2$. In Eq.~(\ref{DI}), 
\begin{equation}
\frac{2k_B T_{sys}}{A_{phy}}\simeq 3.912\times 10^{10}~{\rm Jy} 
\left(\frac{\nu}{100{\rm GHz}}\right)^2 
\left(\frac{T_{sys}}{100{\rm K}}\right) \left(\frac{\lambda}{D}\right)^2.
\end{equation}
With regard to CMB observations, because a given baseline $u$
is sensitive only to a narrow range of $l$ centered at about $2\pi u$,  
it is more suitable to use the sensitivity per baseline 
per polarization given by Eq.~(\ref{DI}) with $N_b=1$ and $N_p=1$,
\begin{equation}
s_b = \frac{1}{\eta_s \eta_a} \frac{2k_B T_{sys}}{A_{phy}}
    \frac{1}{\sqrt{2\Delta_\nu}}.
\end{equation}

Let us consider a simple two-element interferometer with baseline $u$.
For the one in close-packed configuration, it is convenient to 
describe the instrumental noise on a pixelized sky map as in the
single-dish experiment, 
with $\Omega_{pix}=\Omega_A$ and $\sigma_{pix}=s_b/\sqrt{t_{pix}}$,
for which $w^{-1}$ is specified on the celestial sphere.
In the widespread configuration, we can adopt the same strategy but now
on a pixelized u-sphere. The size of the pixel can be chosen as the
resolution in $u$, i.e, $\Omega_{pix} \sim (\Delta u/u)^2$, and the
noise in this pixel is also $\sigma_{pix}=s_b/\sqrt{t_{pix}}$.
Since $l\sim 2\pi u$, we have $\Omega_{pix} \sim (\Delta l_I/l)^2$.
Hence, we can assign $w^{-1}$ on the u-sphere.

\section{Polarization}
\label{poln}

It is a routine for radio interferometers to measure the polarization
of the radiation field~\cite{cot}. Polarization measurements are generally made 
using a pair of feeds on each antenna. Usually, the feeds
are sensitive to orthogonal circular or linear polarizations. For instance,
if the dual-polarization feeds measure the right and left circular
polarizations, then the output will be the four correlations:
$\langle RR^* \rangle$, $\langle RL^* \rangle$,
$\langle LR^* \rangle$, and $\langle LL^* \rangle$.
They can be related to the four Stokes parameters $(I,Q,U,V)$.
Denoting their associated visibilty functions by $(V_I,V_Q,V_U,V_V)$, we have
\begin{eqnarray}
\langle RR^* \rangle &=& V_I + V_V, \nonumber \\
\langle LL^* \rangle &=& V_I - V_V, \nonumber \\
\langle RL^* \rangle &=& V_Q + iV_U \equiv V_+, \nonumber \\
\langle LR^* \rangle &=& V_Q - iV_U \equiv V_-, 
\end{eqnarray}
where we have neglected the parallactic angle of the feed with
respect to the sky and the leakage from one polarization channel to the
other polarization channel. Similiar to Eq.~(\ref{VTb}), we have
\begin{equation}
V_\pm(\vec u) =\int d{\hat e'} A(\hat e',\hat e) (Q\pm iU)(\hat e') 
                             e^{2\pi i{\vec u}\cdot {\hat e'}}.
\label{VQUb}
\end{equation}

It is difficult to analyze these equations in general. We thus
proceed the analysis using the flat-sky approximation,
i.e., $\hat u \cdot \hat e \simeq 0$. 
Let us first set up a rectangular coordinate $(\hat e_x,\hat e_y,\hat e_z)$,
and choose $\hat e=\hat e_x$, then it can be shown that (see Appendix~C)
\begin{eqnarray}
{\partial\!\!'}_{\hat e'}^2 \bar{\partial\!\!'}_{\hat u}^2
e^{2\pi i{\vec u}\cdot {\hat e'}} 
&\simeq& (2\pi u)^4 e^{-2i\theta_u} e^{2\pi i{\vec u}\cdot {\hat e'}},
\label{fsky1}  \\
\bar{\partial\!\!'}_{\hat e'}^2 {\partial\!\!'}_{\hat u}^2
e^{2\pi i{\vec u}\cdot {\hat e'}} 
&\simeq& (2\pi u)^4 e^{2i\theta_u} e^{2\pi i{\vec u}\cdot {\hat e'}},
\label{fsky2}
\end{eqnarray}
where $\theta_u=\hat u \cdot \hat e_y$ is the polar angle of $\vec u$ 
in the $\vec u$-plane parallel to the $\hat e_y-\hat e_z$ plane, and the
basis $(\hat e_y,\hat e_z)$ is used to define $Q$ and $U$.
As such, Eq.~(\ref{VQUb}) becomes
\begin{eqnarray}
V_+(\vec u) &\simeq& (2\pi u)^{-4} e^{2i\theta_u}
\int d{\hat e'} A(\hat e',\hat e) (Q+iU)(\hat e') 
{\partial\!\!'}_{\hat e'}^2 \bar{\partial\!\!'}_{\hat u}^2
e^{2\pi i{\vec u}\cdot {\hat e'}}, \nonumber \\
V_-(\vec u) &\simeq& (2\pi u)^{-4} e^{-2i\theta_u}
\int d{\hat e'} A(\hat e',\hat e) (Q-iU)(\hat e') 
\bar{\partial\!\!'}_{\hat e'}^2 {\partial\!\!'}_{\hat u}^2
e^{2\pi i{\vec u}\cdot {\hat e'}}.                             
\end{eqnarray}
Using Eqs.~(\ref{expand}) and (\ref{bareth}), 
and following the steps to reach Eq.~(\ref{VTW}), we obtain
\begin{eqnarray}
V_\pm(\vec u) \simeq && 8\pi^2 e^{\pm 2i\theta_u} \sum_{l_1 m_1} \sum_{l_2 m_2}
   a_{\mp 2, l_1 m_1} i^{l_2} j_{l_2}(2\pi u) \:_{\mp 2}Y_{l_2 m_2}(\hat u)
   \sqrt{\frac{4\pi}{2l_1+1}}\sqrt{\frac{4\pi}{2l_2+1}} \times \nonumber \\
   && \sum_{m} \:_{-m}Y^*_{l_2 m_2}(\hat e) \:_{-m}Y_{l_1 m_1} (\hat e)
   W^{1\over2}(l_1,l_2,m),   
\label{VQUW}
\end{eqnarray}
where 
\begin{equation}
W^{1\over2}(l_1,l_2,m)=\int_0^\pi 
                d\beta~\beta~e^{-\frac{\beta^2}{2\sigma_b^2}}
                \:_{\mp 2}Y^*_{l_2 m}(\beta,0) \:_{\mp 2}Y_{l_1 m}(\beta,0).
\label{s2winf}
\end{equation}
It has been shown that the window function for polarization measurements
in single-dish experiments is well approximated by that for anisotropy 
as long as $l\gg 2$~\cite{ngliu} (also see Eq.~(\ref{swinf})).
This result can also be applied to Eq.~(\ref{s2winf}).
So, we can approximate the window function~(\ref{s2winf}) by 
Eq.~(\ref{winf}), and hence the subsequent analyses are the same as 
we have treated the anisotropy in the previous sections. 
The only differences are the overall phase factor containing $\theta_u$, 
the spherical harmonics of spin-2, 
and that $\vec u$ lies in the $\vec u$-plane. 

For example, in the widespread configuration, we have 
\begin{equation}
V_\pm(\vec u) \simeq \sqrt{8\pi} \sigma_b e^{\pm 2i\theta_u}
            \sum_{lm} a_{\mp 2,lm} i^l j_l(2\pi u) \:_{\mp 2}Y_{lm}(\hat u).
\label{VQUl2}  
\end{equation}
Analogous to Eq.~(\ref{cfe}), by using Eqs.~(\ref{CMBaa}) and (\ref{add}),
we can construct four independent two-point correlation functions 
from Eqs.~(\ref{VTl2}) and (\ref{VQUl2}). They are 
\begin{eqnarray}
&& \langle V^*_T(\vec u_1) V_T(\vec u_2) \rangle \simeq
    \frac{\sigma_b^2}{8\pi^2} 
    \sum_l (2l+1) C_{Tl} I_l(u) P_l(\cos\theta), \nonumber \\
&& \langle V^*_+(\vec u_1) V_+(\vec u_2) \rangle \simeq
    \frac{\sigma_b^2}{2\pi} e^{-2i\theta}
    \sum_l \sqrt{\frac{2l+1}{4\pi}}
    (C_{El} + C_{Bl}) I_l(u) \:_{2}Y_{l-2}(\theta,0), \nonumber \\
&& \langle V^*_-(\vec u_1) V_+(\vec u_2) \rangle \simeq
    \frac{\sigma_b^2}{2\pi} e^{2i(\theta_{u_1}+\theta_{u_2})}
    \sum_l \sqrt{\frac{2l+1}{4\pi}}
    (C_{El} - C_{Bl}) I_l(u) \:_{2}Y_{l2}(\theta,0), \nonumber \\
&& \langle V^*_T(\vec u_1) V_+(\vec u_2) \rangle \simeq
    \frac{\sigma_b^2}{2\pi} e^{2i\theta_{u_2}}
    \sum_l \frac{2l+1}{4\pi} \sqrt{\frac{(l-2)!}{(l+2)!}} 
    C_{Cl} I_l(u) P_l^2(\cos\theta),
\end{eqnarray}
where $|\vec u_1|=|\vec u_2|=u$,
and the separation angle $\theta =\theta_{u_1}-\theta_{u_2}$.
Hence, the rms total polarization in a given visibility is 
\begin{equation}
V_{P{rms}}^2 \equiv \langle V^*_+(\vec u) V_+(\vec u) \rangle  \simeq
\frac{\sigma_b^2}{8\pi^2} \sum_l (2l+1) (C_{El} + C_{Bl}) I_l(u).
\end{equation}

\section{Estimation of Signal to Noise Ratios}
\label{snr}

In the previous sections, we have presented the basic results that can allow us
to lay out the strategy in CMB interferometric observation, and to 
deal with the observed data. First of all, it is useful to make some simple
estimates of the CMB signal to noise ratios for the upcoming CMB 
interferometers.   

For a close-packed interferometer, the CMB anisotropy signal in a given 
visibility is given by 
$S_T= (\partial B_\nu/\partial T) V_{T{rms}}/\sqrt{2}$, 
where from Eq.~(\ref{cpv2}),
\begin{equation}
V_{T{rms}} \equiv \langle V_T^*(\hat e) V_T(\hat e) \rangle^{1\over 2}
\simeq \frac{1}{\sqrt 2} \frac{\sigma_b}{l_{min}}
 \left[ \sum_l (2l+1) C_{Tl} e^{-(l-l_{min})^2 \sigma_b^2} \right]^{1\over2},
\end{equation}
while the noise limit $N$ is given by $\Delta I$~(\ref{DI}). 
Hence, we estimate the signal to noise ratio per single pointing as
\begin{eqnarray}
\frac{S_T}{N}
&\simeq& 1.6 \eta_s \sqrt\eta_a \sqrt{N_b N_p} f(x)
 \left(\frac{\Delta_\nu}{10{\rm GHz}}\right)^{1\over 2}
 \left(\frac{t_{int}}{\rm hr}\right)^{1\over 2}
 \left(\frac{100{\rm GHz}}{\nu}\right)
 \left(\frac{100{\rm K}}{T_{sys}}\right) \left(\frac{D}{1 {\rm m}}\right)
 \nonumber \\
 && \times \left[\frac{1}{(\mu{\rm K})^2} \frac{1}{l_{min}^2}
     \sum_l (2l+1) C_{Tl} e^{-(l-l_{min})^2 \sigma_b^2} \right]^{1\over 2},
\label{STN}
\end{eqnarray}
where $f(x)$ is given by Eq.~(\ref{f(x)}). 
This formula is also applied for the CMB polarization except replacing
$C_{Tl}$ by $C_{El}+C_{Bl}$.

To estimate the S/N ratios, we simply neglect the B-polarization power spectrum, 
and approximate the anisotropy and E-polarization spectra for $l=300-1000$ 
respectively by
\begin{eqnarray}
l(l+1) C_{Tl} &\simeq& 2\pi (\Delta T)^2,\quad\quad 
   \Delta T \simeq 50 \mu{\rm K}; \nonumber \\
l(l+1) C_{El} &\simeq& 2\pi (\Delta E)^2,\quad\quad 
   \Delta E \simeq 5 \mu{\rm K}.
\end{eqnarray}
Let us consider a close-packed interferometer with 19 dishes 
in the hexagonal configuration. Then, it has $171$ baselines in total,
and the number of shortest baselines is $42$.
For $D=1.2{\rm m}$, $\nu=30{\rm GHz}$, $N_b=42$, and $N_p=2$, 
the minimum spacing $l_{min}=2\pi u\simeq 2\pi D/\lambda \simeq 754$. 
Hence the anisotropy and E-polarization S/N ratios are respectively given by
\begin{eqnarray}
{S_T\over N} &\simeq& 3 \eta_s \sqrt\eta_a 
 \left(\frac{\Delta_\nu}{10{\rm GHz}}\right)^{1\over 2}
 \left(\frac{t_{int}}{\rm hr}\right)^{1\over 2}
 \left(\frac{100{\rm K}}{T_{sys}}\right) 
 \left(\frac{\Delta T}{50\mu{\rm K}}\right), \nonumber \\
{S_E\over N} &\simeq& 0.3 \eta_s \sqrt\eta_a 
 \left(\frac{\Delta_\nu}{10{\rm GHz}}\right)^{1\over 2}
 \left(\frac{t_{int}}{\rm hr}\right)^{1\over 2}
 \left(\frac{100{\rm K}}{T_{sys}}\right) 
 \left(\frac{\Delta E}{5\mu{\rm K}}\right).
\end{eqnarray}
If we switch to a higher frequency $\nu=90{\rm GHz}$ while fixing
$l_{min}\simeq 754$, then the dish size should be reduced to $D=0.4{\rm m}$.
For $N_b=42$ and $N_p=2$, we would expect the same S/N ratios as 
for $\nu=30{\rm GHz}$.
However, corrected for the Rayleigh-Jeans limit, we find that
the prefactor in the $S_T/N$ is reduced to $2.5$,
whereas in the $S_E/N$ the prefactor is $0.25$.

\section{Conclusions and Discussions}
\label{con}

We have presented a full-sky analysis of the monochromatic CMB 
complex visibilities. First of all, an exact expression for
the sky power spectrum is obtained in Eq.~(\ref{power}). It has an advantage
over the flat-sky approximation for the interference function 
being positively definite, whereas the flat-sky interference function 
is rapidly oscillating about the zero. 
In the latter, care must be taken in summing over $l$ 
for a flat spectrum due to significant cancellations~\cite{white}. 
Moreover, we have found that the flat-sky approximation generally
underestimates the power spectrum.

A full-sky expression in Eq.~(\ref{VTW}) for the CMB complex visibility is 
our main result. It serves as the basis for analyzing the 
$l$-resolution in a given visibility, especially when a large sky 
scanning is needed in order to obtain a high resolution in $l$-space. 
We have shown in Eq.~(\ref{resol})
that the full-sky scanning can increase the $l$-resolution
from maximal $\Delta l \sim D/\lambda$ to $\Delta l \sim l^{1\over 3}$. 
One should further check whether the resolution increases linearly with the
sky coverage.

We have worked out two limiting cases of the visibility equation~(\ref{VTW}) 
in which the statistics of the visibility becomes transparent. 
Firstly, we have shown that the close-packed interferometer is
functioning like a single-dish switching experiment. 
Therefore, on the issue of obtaining hign $l$-resolution, it is important to 
study and compare the efficiency of the the usual method of synthesis imaging 
of the sky against that of the aforementioned sky scanning method.
Secondly, we have suggested for the widespread configuration that 
one should analyse the visibility data on the u-sphere. 
We have also pointed out that the interferometry can in principle avoid
the cosmic variance by obtaining different u-spheres via multiple pointings of 
the telescope to uncorrelated patches of the CMB sky. It is interesting 
to study how to implement this concept in practical situations.

In this paper, we have performed the calculations assuming a monochromatic
electromagnetic source, and allowing a nonvanishing geometrical delay 
$\tau = \lambda\vec u\cdot \hat e_0$,
which measures the elapsed time for the wavefront reaching one antenna 
and then the other. However, when observing with a finite bandwidth 
$\Delta_\nu$, one usually correlates signals at two separate points 
on the same wavefront in order to obtain full fringe. This can be done
by including within the interferometric system a computer-controlled
phase delay to compensate for $\tau$~\cite{thom}. 
As a consequence, the interferometer response is
\begin{equation}
V(\vec u, \tau) = \int_{\nu_0-\Delta_\nu/2}^{\nu_0+\Delta_\nu/2}
V(\vec u, \nu) e^{2\pi i \tau \nu} d\nu.
\end{equation}
This is a Fourier transform with conjugate variables $\nu$ and $\tau$, and
can be inverted to extract the desired $V(\vec u, \nu)$.

Finally, we remark that CMB interferometric observations are radically
different from traditional radio interferometry. 
We certainly need more studies on
several important issues such as observing strategy, $l$-space resolution and 
mosaicing, optimal estimation of the power spectra, 
point source and other foreground subtraction, and ground pickup removal.

\section*{Acknowledgments}

The author would like to thank D. Bond, T.-H. Chiueh, H. Liang, J. Lim,
K.-Y. Lo, U.-L. Pen, S. Prunet, and R. Subrahmanyan for their useful 
discussions. He is also grateful to CITA for their hospitality during 
his sabbatical year, where most of the work has been done.
This work was supported in part by the National Science Council, ROC 
under the Grant NSC89-2112-M-001-001, and NSC87-37047F.

\section*{Appendix~A: Spin-weighted Spherical Harmonics}

The spin-weighted spherical harmonics are related to the representation 
matrices of the three-dimensional rotation group. 
If we define a rotation $R(\alpha,\beta,\gamma)$ as being 
composed of a rotation $\alpha$ around $\hat e_z$, followed by $\beta$ around
the new $\hat e_y'$ and finally $\gamma$ around $\hat e_z''$,
the rotation matrix of $R$ will be given by~\cite{new}
\begin{equation}
D_{-sm}^{l}(\alpha,\beta,\gamma)=\sqrt{\frac{4\pi}{2l+1}}
\:_{s}Y_{lm}(\beta,\alpha)\;e^{-is\gamma}.
\end{equation}

An explicit expression of the spin-$s$ spherical harmonics is~\footnote
{In Ref.~\cite{new}, the sign $(-1)^m$ is absent. 
We have added the sign in order to match the conventional 
definition for $Y_{lm}$.}
\cite{new,pen}
\begin{eqnarray}
\:_{s}Y_{lm}(\theta,\phi)&&=(-1)^{m}e^{im\phi}\left[\frac{2l+1}{4\pi}
\frac{(l+m)!}{(l+s)!}\frac{(l-m)!}{(l-s)!}\right]^{\frac{1}{2}}
\sin^{2l} \left(\frac{\theta}{2}\right) \nonumber \\
&&\times\sum_{r}\left(\begin{array}{c}
l-s\\
r\\
\end{array}\right)\left(\begin{array}{c}
l+s\\
r+s-m\\
\end{array}\right)(-1)^{l-s-r}\cot^{2r+s-m}\left(\frac{\theta}{2}\right),
\label{s-har}
\end{eqnarray}
where
\begin{equation}
\max(0,m-s) \le r\le \min(l-s,l+m). \nonumber
\end{equation}
Note that the ordinary spherical harmonics $Y_{lm}=\:_{0}Y_{lm}$. 
Using the expression~(\ref{s-har}), one can show the symmetry,
\begin{equation}
D_{-sm}^{l}(\alpha,\beta,\gamma)=D_{m-s}^{l}(\gamma,\beta,\alpha).
\label{sym}
\end{equation}
They have the conjugation relation and parity relation,
\begin{equation}
\:_{s}Y^*_{lm}(\theta,\phi)=(-1)^{m+s}\:_{-s}Y_{l-m}(\theta,\phi),
\end{equation}
\begin{equation}
\:_{s}Y_{lm}(\pi-\theta,\phi+\pi)=(-1)^{l}\:_{-s}Y_{lm}(\theta,\phi).
\end{equation}
They satisfy the orthonormality condition and completeness relation,
\begin{equation}
\int d\Omega \:_{s}Y^*_{l'm'}(\theta,\phi)\:_{s}Y_{lm}(\theta,\phi)
=\delta_{l'l}\delta_{m'm},
\label{ortho}
\end{equation}
\begin{equation}
\sum_{lm}\:_{s}Y^*_{lm}(\theta',\phi')\:_{s}Y_{lm}(\theta,\phi)
=\delta(\phi'-\phi) \delta(\cos\theta'-\cos\theta).
\end{equation}
Therefore, a quantity $\eta$ of spin-weight $s$ defined on the sphere 
can be expanded in spin-$s$ basis,
\begin{equation}
\eta(\theta,\phi)=\sum_{lm} \eta_{lm} \:_{s}Y_{lm} (\theta,\phi),
\end{equation}
where the expansion coefficients $\eta_{lm}$ are scalars. 

The raising and lowering operators, $\partial\!\!'$ and 
$\bar{\partial\!\!'}$,
acting on $\eta$ of spin-weight $s$, are defined by~\cite{new}
\begin{eqnarray}
{\partial\!\!'}\eta&=&-(\sin\theta)^s \left[\frac{\partial}{\partial\theta}
  +i\csc\theta\frac{\partial}{\partial\phi}\right](\sin\theta)^{-s}\eta,
\nonumber \\
\bar{\partial\!\!'}\eta&=&-(\sin\theta)^{-s}\left[\frac{\partial}{\partial
  \theta} -i\csc\theta\frac{\partial}{\partial\phi}\right](\sin\theta)^s\eta.
\end{eqnarray}
When they act on the spin-$s$ spherical harmonics, we have~\cite{new}
\begin{eqnarray}
{\partial\!\!'}\:_{s}Y_{lm}&=&\left[(l-s)(l+s+1)\right]^{1\over 2} 
                            \:_{s+1}Y_{lm}, 
\nonumber \\
\bar{\partial\!\!'}\:_{s}Y_{lm}&=&-\left[(l+s)(l-s+1)\right]^{1\over 2} 
                            \:_{s-1}Y_{lm}. 
\label{bareth}
\end{eqnarray}
Using these raising and lowering operations, one can obtain the generalized 
recursion relation~\cite{ngliu}, which allows one to construct easily
the high-$l$ spin-weighted harmonics from the low-$l$ harmonics.

From the rotation group multiplication law, one can derive
the generalized addition theorem~\cite{ngliu},
\begin{equation}
\sum_{m}\:_{s_1}Y^*_{lm}(\theta_1,\phi_1)
          \:_{s_2}Y_{lm}(\theta_2,\phi_2) 
        =\sqrt{\frac{2l+1}{4\pi}}(-1)^{s_1-s_2}
           \:_{-s_1}Y_{ls_2}(\beta,\alpha_2)\;e^{-is_1\alpha_1},
\label{add}
\end{equation}
where $\beta$ is the separation angle between the two pointings
$(\theta_1,\phi_1)$ and $(\theta_2,\phi_2)$ on the celestial sphere.
Connecting them by a geodesic, $\alpha_1$ and $\alpha_2$ are 
the angles between the geodesic and the longitudes passing through 
$(\theta_1,\phi_1)$ and $(\theta_2,\phi_2)$ respectively. 
When $s_1=s_2=0$, Eq.~(\ref{add}) reduces to the familiar addition
theorem for spherical harmonics. Similarly, we have
\begin{equation}
\:_{s}Y_{lm}(\theta_1,\phi_1)=\sqrt{\frac{4\pi}{2l+1}}
\sum_{m'}\:_{s}Y_{lm'}(\beta,\alpha_2)\,e^{is\alpha_1}
         \:_{-m'}Y_{lm}(\theta_2,\phi_2).
\label{add2}
\end{equation}
Furthermore, using the symmetry relation~(\ref{sym}) and 
the addition theorem~(\ref{add}), we obtain
\begin{equation}
\sum_{m}\:_{-m}Y^*_{l-s_1}(\theta_1,\phi_1)
          \:_{-m}Y_{l-s_2}(\theta_2,\phi_2) 
        =\sqrt{\frac{2l+1}{4\pi}}(-1)^{s_1-s_2}
           \:_{-s_1}Y_{ls_2}(\theta_2-\theta_1,0)\;
           e^{is_1\phi_1} e^{-is_2\phi_2}.
\end{equation}
As long as the two pointings are identical, this becomes
\begin{equation}
\sum_{m}\:_{-m}Y^*_{l-s_1}(\theta,\phi)
          \:_{-m}Y_{l-s_2}(\theta,\phi) 
        =\frac{2l+1}{4\pi} \delta_{s_1 s_2}.
\label{yy}
\end{equation}

\section*{Appendix~B: Window Function}

The window function $W^{1\over2}(l_1,l_2,m)$ in Eq.~(\ref{winf}) is 
\begin{eqnarray}
W^{1\over2} 
&=& \int_0^\pi d\beta~\beta~e^{-\frac{\beta^2}{2\sigma_b^2}}
    Y^*_{l_2 m}(\beta,0) Y_{l_1 m}(\beta,0) \nonumber \\
&=& \left[\frac{2l_1+1}{4\pi} \frac{(l_1-m)!}{(l_1+m)!} \right]^{1\over2}
    \left[\frac{2l_2+1}{4\pi} \frac{(l_2-m)!}{(l_2+m)!} \right]^{1\over2}~I,
\end{eqnarray}
where
\begin{equation}    
I= \int_0^\pi d\beta~\beta~e^{-\frac{\beta^2}{2\sigma_b^2}}
    P_{l_1}^m(\cos\beta) P_{l_2}^m(\cos\beta),
\label{I}
\end{equation}
and we restrict $m\ge 0$. This has already included the case with $m<0$, 
because $W^{1\over2}(l_1,l_2,m)= W^{1\over2}(l_1,l_2,-m)$.

Since the Gaussian function in the integral~(\ref{I}) has a width of 
$\sigma_b \sim l_{min}^{-1}$, we approximate the associate Legendre 
polynomial as~\cite{gra}
\begin{eqnarray}
&&P_l^m(\cos\beta) \sim \sqrt{\frac{2}{\pi}}
  \frac{\Gamma(l+m+1)}{\Gamma(l+{3\over2})} \sin^{-{1\over2}}\beta~
  \cos\left[(l+{1\over2})\beta + (m-{1\over2}){\pi\over2}\right] 
  \quad{\rm for}\quad  l \gg l_{min}, \nonumber \\
&&P_l^{-m}(\cos\beta) \sim l^{-m} J_m(l\beta)
  \quad{\rm for}\quad  l \sim l_{min}, \nonumber \\
&&P_l^m(\cos\beta) \sim (-1)^m \frac{(l+m)!}{m!(l-m)!} 
  \left({\beta\over2}\right)^m
  \quad{\rm for}\quad  l \ll l_{min}. 
\end{eqnarray}
Then, the integral can be integrated analytically for certain limiting
values of $l_1$ and $l_2$.

For $l_1 \sim l_{min}$ and $l_2 \sim l_{min}$,
\begin{eqnarray}
I &=& 
\frac{(l_1+m)!}{(l_1-m)!}\frac{(l_2+m)!}{(l_2-m)!} 
\int_0^\pi d\beta~\beta~e^{-\frac{\beta^2}{2\sigma_b^2}}
    P_{l_1}^{-m}(\cos\beta) P_{l_2}^{-m}(\cos\beta) \nonumber \\
&\simeq& 
\frac{(l_1+m)!}{(l_1-m)!}\frac{(l_2+m)!}{(l_2-m)!}\frac{1}{l_1^m l_2^m}
\int_0^\infty d\beta~\beta~e^{-\frac{\beta^2}{2\sigma_b^2}} 
J_m(l_1\beta) J_m(l_2\beta) \nonumber \\
&=& \frac{(l_1+m)!}{(l_1-m)!}\frac{(l_2+m)!}{(l_2-m)!}
    \frac{\sigma_b^2}{l_1^m l_2^m} e^{-{1\over2}(l_1^2+l_2^2)\sigma_b^2}
    I_m(l_1 l_2 \sigma_b^2),
\end{eqnarray}
where $I_m$ is the modified Bessel function, which has the limiting form
\begin{equation}
I_m(x) \sim \frac{1}{\sqrt{2\pi x}}~e^x \quad{\rm for}\quad x\gg 1.
\end{equation}
When $l_1 l_2 \sigma_b^2 \gg 1$, we have
\begin{equation}
I\simeq \frac{(l_1+m)!}{(l_1-m)!}\frac{(l_2+m)!}{(l_2-m)!}
  \frac{1}{(l_1 l_2)^{m+{1\over2}}} \frac{\sigma_b}{\sqrt{2\pi}}
  e^{-{1\over2}(l_1-l_2)^2 \sigma_b^2}.
\end{equation}
Hence this gives the result in Eq.~(\ref{cpwf}).

For $l_1 \gg l_{min}$ and $l_1 \gg l_{min}$,
\begin{eqnarray}
I &\simeq& {2\over\pi}
\frac{\Gamma(l_1+m+1)}{\Gamma(l_1+{3\over2})}
\frac{\Gamma(l_2+m+1)}{\Gamma(l_2+{3\over2})}
\int_0^\pi d\beta~\beta~e^{-\frac{\beta^2}{2\sigma_b^2}} \sin^{-1}\beta 
\times \nonumber \\
&&\cos\left[(l_1+{1\over2})\beta + (m-{1\over2}){\pi\over2}\right] 
\cos\left[(l_2+{1\over2})\beta + (m-{1\over2}){\pi\over2}\right]  \nonumber \\
&\simeq& {1\over\pi}
\frac{\Gamma(l_1+m+1)}{\Gamma(l_1+{3\over2})}
\frac{\Gamma(l_2+m+1)}{\Gamma(l_2+{3\over2})}
\int_0^\infty d\beta~e^{-\frac{\beta^2}{2\sigma_b^2}} 
\cos\left[(l_1-l_2)\beta\right]  \nonumber \\
&=& 
\frac{\Gamma(l_1+m+1)}{\Gamma(l_1+{3\over2})}
\frac{\Gamma(l_2+m+1)}{\Gamma(l_2+{3\over2})}
\frac{\sigma_b}{\sqrt{2\pi}} e^{-{1\over2}(l_1-l_2)^2 \sigma_b^2}.
\end{eqnarray}
Hence this gives the result in Eq.~(\ref{wswf}).

Furthermore, we have found that when $l_2\gg l_{min}$,
$I$ is subdominant for $l_1 \le l_{min}$. 
When $l_2 \sim l_{min}$, $I$ is subdominant for both $l_1 \gg l_{min}$ and 
$l_1 \ll l_{min}$.

\section*{Appendix~C: Flat-sky Approximation}

We are going to evaluate 
\begin{equation}
{\partial\!\!'}_2^2 \bar{\partial\!\!'}_1^2
e^{2\pi i u {\hat e_1}\cdot {\hat e_2}}, \nonumber 
\end{equation}
under the condition that $\hat e_1 \cdot \hat e_2 \simeq 0$. 
This condition corresponds to the flat-sky approximation when 
$\vec u=u\hat e_1$ is the baseline vector and $\hat e_2$ is the telescope 
pointing direction. The reader may refer to Refs.~\cite{zald,kamion,white}
for different approaches.

Using Eqs.~(\ref{expon}), (\ref{bareth}), and (\ref{add}), we obtain
\begin{eqnarray}
\bar{\partial\!\!'}_1^2 e^{2\pi i u {\hat e_1}\cdot {\hat e_2}}
&=& 4\pi\sum_l i^l j_l(2\pi u) 
  \left[\frac{2l+1}{4\pi}\frac{(l+2)!}{(l-2)!}\right]^{1\over2}
  Y_{l -2} (\hat e_1 \cdot \hat e_2,\alpha_1) \nonumber \\
&=& -4\pi\sum_l i^l j_l(2\pi u) 
  \frac{2l+1}{4\pi} \frac{(l+2)!}{(l-2)!} P_l^{-2} (\hat e_1 \cdot \hat e_2),
\label{ethexp}
\end{eqnarray}
where we have substituted $\alpha_1\simeq \pi/2$ 
under the flat-sky approximation.
From the recursion relation,  
\begin{equation}
P_l^{m+2}(x)+2(m+1)\frac{x}{\sqrt{1-x^2}}P_l^{m+1}(x)+(l-m)(l+m+1)P_l^m(x)=0,
\end{equation}
when $x\simeq 0$ and $m=-2$, we have 
\begin{equation}
P_l(x)\simeq -(l+2)(l-1) P_l^{-2}(x).
\label{pl02}
\end{equation}
As such, Eq.~(\ref{ethexp}) can be approximated as
\begin{equation}
\bar{\partial\!\!'}_1^2 e^{2\pi i u {\hat e_1}\cdot {\hat e_2}} \simeq 
4\pi \sum_{lm} i^l j_l(2\pi u) l(l+1) Y_{lm}^*(\hat e_1) Y_{lm}(\hat e_2).
\label{ethexp2}
\end{equation}
Applying the operator ${\partial\!\!'}_2^2$ to Eq.~(\ref{ethexp2}) and
doing the approximation~(\ref{pl02}) again, we find that 
\begin{equation}
{\partial\!\!'}_2^2 \bar{\partial\!\!'}_1^2 
e^{2\pi i u {\hat e_1}\cdot {\hat e_2}}
\simeq  e^{-2i\theta_u}
\sum_l i^l j_l(2\pi u) l^2(l+1)^2(2l+1) P_l(\hat e_1 \cdot \hat e_2),
\label{ethexp3}
\end{equation}
where $\theta_u=\pi/2-\alpha_2$. 
Since $j_l$ is a sharply peaked function at $l\sim 2\pi u$ for $2\pi u\gg 1$,  
we can approximate the factor $l^2(l+1)^2$ in Eq.~(\ref{ethexp3}) 
by $(2\pi u)^4$ and then take it out of the $l$-summation. Hence we obtain
Eq.~(\ref{fsky1}). We can follow the same steps to derive Eq.~(\ref{fsky2}).

\newpage

\begin{figure}
\leavevmode
\hbox{
\epsfxsize=6.5in
\epsffile{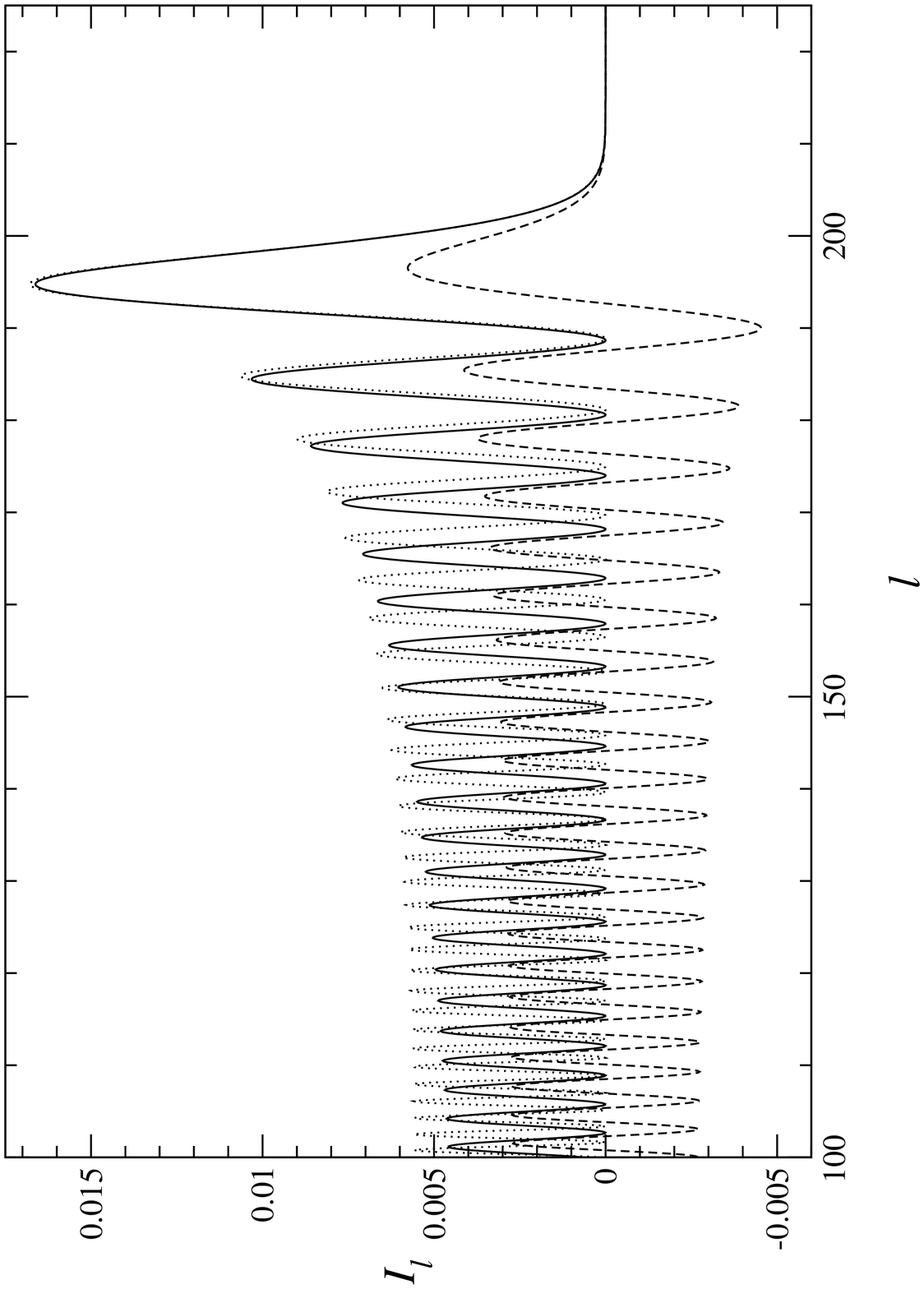}}
\caption{}
\label{fig1}
\end{figure}

\begin{center}
{\bf FIGURE CAPTIONS}
\end{center}
\bigskip
\noindent
Fig.~\ref{fig1}  
Solid and dotted curves are the interference functions $I_l(u)$ 
with $2\pi u=200$ plotted using Eq.~(\ref{intf}) and the asymptotic 
expansion~(\ref{asymj}) respectively. Dashed curve is the flat-sky 
approximation in Eq.~(\ref{fsintf}).
\newpage

\newpage

\end{document}